\documentclass{jfm}
\usepackage{graphicx}
\usepackage{amsmath}
\usepackage{epstopdf, epsfig}
\usepackage{xcolor}
\usepackage{color,soul}
\usepackage{upgreek}
\usepackage{units}
\usepackage{stmaryrd}
\shorttitle{Active Jurin's law}
\shortauthor{B. Mandal and J. Chaudhuri}

\title{Active Jurin's law}

\author{Birendra Mandal
        and Joydip Chaudhuri
    \corresp{\email{joy@che.iitkgp.ac.in}}}
\affiliation{Department of Chemical Engineering, Indian Institute of Technology Kharagpur, Kharagpur 721302, West Bengal, India}

\begin{document}
\maketitle

\begin{abstract}

Capillary rise is one of the classical problems in fluid mechanics and is traditionally described by Jurin's law, which balances capillary suction against hydrostatic pressure. 
Here we extend this classical result to active fluids, materials that generate internal stresses through microscopic energy consumption. 
Using the continuum theory of active nematics, we show that activity modifies the normal stress balance at the liquid–gas interface through an additional active normal stress contribution. 
This leads to a generalized \emph{active Jurin's law}, which can be written in dimensionless form as, $H_{\infty} = 1 - \mathrm{Ja}_a \xi_0$, where, $H_{\infty}$ is the dimensionless active Jurin height at equilibrium, $\mathrm{Ja}_a$ is an active Jurin number comparing active stress to capillary pressure, and $\xi_0$ characterizes the alignment of active constituents at the meniscus. 
The theory predicts that extensile and contractile active fluids can either enhance or suppress capillary rise depending on the magnitude of activity and the interfacial alignment state. 
From this relation we construct a phase diagram in the $(\mathrm{Ja}_a,\xi_0)$ plane that delineates regimes of activity-enhanced rise, activity-suppressed rise, and complete suppression of the classical capillary state. 
When orientational order depends on confinement and flow, the coupling between activity and capillarity produces nonlinear equilibrium conditions that may admit multiple steady heights; linear stability analysis (LSA) reveals that the overdamped dynamics selects a single stable state, whereas the inertial extension allows the possibility of \emph{activity-induced bistability}. 
These results show that internally generated stresses fundamentally reshape one of the most classical capillary transport problems.

\end{abstract}

\section{Introduction}

Capillary rise is one of the most classical problems in fluid mechanics and arises from the balance between capillary suction at a curved liquid–gas interface and hydrostatic pressure in the liquid column \citep{weislogel2012compound, kim2017capillary, zhou2020universality}. 
For a passive Newtonian liquid in a cylindrical capillary tube of radius $a$, the equilibrium rise height ($h_J$) is determined by Jurin's law \citep{jurin1718ii},
\begin{equation}
h_J = \frac{2\gamma \cos\theta}{\rho g a},
\label{eq:jurin}
\end{equation}
where, the symbols, $\gamma$, $\theta$, $\rho$, and $g$ denote the surface tension, contact angle, liquid density and gravitational acceleration, respectively. This classical relation represents a simple mechanical balance between Laplace pressure and hydrostatic weight, and plays a central role in problems ranging from porous media transport \citep{parmigiani2011pore,chauhan2021dynamics} and soil mechanics \citep{lu2004rate} to microfluidics \citep{olanrewaju2018capillary} and biological flows \citep{chakraborty2005dynamics}.

In recent years, increasing attention has been directed toward \emph{active fluids}, a class of materials whose microscopic constituents continuously consume energy and generate mechanical stresses internally \citep{cates2018theories,wysocki2020capillary,chisholm2021driven,freund2023object,das2026capillary}. 
Examples include suspensions of swimming bacteria, cytoskeletal extracts containing molecular motors, and active liquid crystals. 
At the continuum level, these systems are commonly described using active nematic hydrodynamics, in which an additional \emph{active stress} arises from the collective orientation of the active particles \citep{doostmohammadi2018active}. 
Unlike passive stresses, this internally generated stress can drive spontaneous flows and modify mechanical balances even in the absence of external forcing.

Since capillary rise is determined by the normal stress balance at the liquid interface, the presence of active stress is expected to alter the capillary pressure that drives imbibition. 
In particular, active stresses contribute an additional normal stress at the meniscus, which can either enhance or oppose the classical Laplace pressure depending on the nature of the activity and the orientation of the active constituents. 
Despite the ubiquity of active stresses in biological and synthetic active materials, their influence on one of the most fundamental capillary phenomena, Jurin's law, has received little attention \citep{liu2018jurin,thiele2026active}.

In this work, we develop a theoretical framework for capillary rise in active fluids. 
Using the continuum description of active nematics, we derive a generalized \emph{active Jurin's law} relating the equilibrium height to the magnitude of activity and the alignment of active constituents at the interface. 
The result can be written in dimensionless form as, $\displaystyle H_{\infty} = 1 - \mathrm{Ja}_a \xi_0 $, where, $\mathrm{Ja}_a$ is an \emph{active Jurin number} comparing active stress to capillary pressure and $\xi_0$ characterizes the interfacial alignment of the active particles. 
This relation shows that activity can either enhance or suppress capillary rise depending on the sign and magnitude of the active stress. The theory further predicts a phase diagram in the $(\mathrm{Ja}_a,\xi_0)$ plane separating regimes of activity-enhanced rise, activity-reduced rise and complete suppression of the classical capillary state. 
When the orientational order depends on confinement or flow, the coupling between activity and capillarity can introduce nonlinear feedback into the meniscus stress balance, leading to multiple steady heights and activity-induced bistability. 
For typical active stresses observed in bacterial suspensions and cytoskeletal extracts, the active Jurin number can approach unity in microcapillaries of radius $10$--$100~\mu$m, suggesting that these effects should be experimentally observable.

The present analysis therefore shows that internally generated stresses fundamentally modify the classical capillary rise problem. 
Beyond a simple renormalization of Jurin's law, activity introduces new regimes and nonlinear behaviours that have no passive analogue, highlighting how active stresses reshape even the most elementary capillary transport processes.

\section{Mathematical formulation}

We consider an incompressible active fluid rising in a vertical cylindrical capillary tube of radius $a$. The height of the liquid column above the reservoir level is denoted by $h(t)$.

The stress tensor of an active nematic fluid is written as \citep{doostmohammadi2018active,alert2020universal},
\begin{equation}
\boldsymbol{\sigma}
=
-p\mathbf{I}
+
2\eta \mathbf{E}
+
\boldsymbol{\sigma}^{\mathrm{act}},
\end{equation}
where $p$ is the pressure, $\eta$ the viscosity and $\mathbf{E}$ the rate-of-strain tensor.

The active stress takes the form \citep{aditi2002hydrodynamic,ramaswamy2010mechanics,Joanny_Ramaswamy_2012,shankar2022optimal},
\begin{equation}\label{active stress 1}
\boldsymbol{\sigma}^{\mathrm{act}} = -\zeta \mathbf{Q},
\end{equation}
where, $\zeta$, is the activity coefficient, $\mathbf{Q}$ is the nematic order tensor and can be expressed as \citep{aditi2002hydrodynamic,ramaswamy2010mechanics,mondal2018electric,shankar2022optimal},
\begin{equation}\label{active stress 2}
\mathbf{Q} = S\left(\mathbf{p}\mathbf{p} - \frac{1}{3}\mathbf{I}\right).
\end{equation}

The symbol, $S$ denotes the scalar order parameter and $\mathbf{p}$ is the director describing the average orientation of the active constituents. The sign of $\zeta$ distinguishes two classes of active fluids. The activity coefficient, $\zeta > 0$ ($\zeta < 0$) describes extensile (contractile) systems \citep{ramaswamy2010mechanics}.

At the liquid–gas interface, the normal stress balance can be expressed as, $\displaystyle \mathbf{n}\cdot(\boldsymbol{\sigma}^{\ell}-\boldsymbol{\sigma}^{g})\cdot\mathbf{n} = \gamma \kappa,$ where, $\mathbf{n}$ is the unit normal and $\kappa$ is the curvature of the interface. Assuming the gas pressure is uniform, the pressure jump across the interface becomes, $\displaystyle p_\ell - p_g = \gamma \kappa + \mathbf{n}\cdot\boldsymbol{\sigma}^{act}\cdot\mathbf{n}.$ For a cylindrical capillary, the curvature ($\kappa$) can be written as, $\displaystyle \kappa \approx {2\cos\theta}/{a}.$ Using the active stress expressions in equations~(\ref{active stress 1}) and (\ref{active stress 2}),
\begin{equation}
\mathbf{n}\cdot\boldsymbol{\sigma}^{act}\cdot\mathbf{n} = 
-\zeta \mathbf{n}\cdot\mathbf{Q}\cdot\mathbf{n} = -\zeta S\left[(\mathbf{p}\cdot\mathbf{n})^2 - \frac13\right].
\end{equation}

Therefore the modified Young-Laplace equation can be expressed as,
\begin{equation}
p_\ell - p_g
=
\frac{2\gamma \cos\theta}{a}
-
\zeta S\left[(\mathbf{p}\cdot\mathbf{n})^2 - \frac13\right].
\end{equation}

Furthermore, hydrostatic balance within the liquid column gives, $\displaystyle p_\ell - p_g = \rho g h.$ Combining the above relations yields the equilibrium rise height,
\begin{equation}\label{gen_Jurin_law}
h_{\infty}^a
=
\frac{1}{\rho g}
\left[
\frac{2\gamma \cos\theta}{a}
-
\zeta S\left((\mathbf{p}\cdot\mathbf{n})^2-\frac13\right)
\right].
\end{equation}

The equation~(\ref{gen_Jurin_law}) represents a generalized Jurin's law for active fluids. The second term corresponds to the normal stress contribution from activity. It can also be interpreted as an effective active surface tension ($\gamma_{\mathrm{eff}}^a$), which can be expressed as, $\displaystyle \gamma_{\mathrm{eff}}^a
=
\gamma
-
\frac{a}{2\cos\theta}
\zeta S
\left[(\mathbf{p}\cdot\mathbf{n})^2-\frac13\right].
$ The dynamics of capillary rise are governed by the balance between viscous resistance and pressure driving. Extending the classical Lucas--Washburn (LW) equation \citep{hamraoui2002analytical} for the active fluids gives rise to the modified LW equation as,
\begin{equation}\label{eq:mod_LW}
\frac{8\eta}{a^2} h \dot{h}
=
\frac{2\gamma \cos\theta}{a}
-
\zeta S\left[(\mathbf{p}\cdot\mathbf{n})^2 - \frac13\right]
-
\rho g h .
\end{equation}

Equation~(\ref{eq:mod_LW}) assumes that the scalar order parameter $S$ and the interfacial alignment remain constant. In realistic active fluids, however, both the degree of orientational order and the alignment of active constituents near the interface may depend on confinement and flow \citep{ramaswamy2010mechanics,marchetti2013hydrodynamics,shankar2022optimal}. To describe this effect, we introduce the interfacial alignment factor, $\displaystyle \xi(h,\dot h) = (\mathbf p\cdot \mathbf n)^2-1/3,$ and allow both $S$ and $\xi$ to depend on the instantaneous height ($h$) and rise velocity ($\dot h$),
\begin{equation}
S=S(h,\dot h),
\qquad
\xi=\xi(h,\dot h).
\end{equation}

The dynamic capillary rise therefore takes the following modified LW form as,
\begin{equation}
\frac{8\eta}{a^2}h\dot h
=
\frac{2\gamma\cos\theta}{a}
-
\zeta S(h,\dot h)\xi(h,\dot h)
-
\rho g h .
\label{eq:dynamic_general}
\end{equation}

It is interesting to note that, although the present theory is formulated for active fluids, the resulting stress balance offers an instructive analogy with long-distance water transport in plants. In a passive capillary, Jurin’s law limits the rise height to $O(1\,\mathrm{m})$ for xylem-sized conduits. Yet in tall trees water is transported to heights of order $10^2\,\mathrm{m}$ due to the negative pressure generated by transpiration at the leaves \citep{tyree2002cohesion,brown2025trees}. From a mechanical viewpoint, this transpiration-induced tension provides an additional driving stress that augments classical capillary suction. In the present formulation, the activity-induced normal stress $(-\zeta S\xi)$ plays an analogous mathematical role by modifying the effective pressure at the meniscus. Although the physical origins differ, both systems may therefore be viewed within a common framework where the column height results from the competition between gravity and an additional driving stress beyond classical capillarity.

\subsection{Linear-response model}\label{linear_response_model}

To obtain an analytical description, we assume weak dependence of $S$ and $\xi$ on $h$ and $\dot h$,
\begin{align}
S(h,\dot h)
&=
S_0+\alpha h+\beta\dot h,\\
\xi(h,\dot h)
&=
\xi_0+\mu h+\nu\dot h.
\end{align}

Substituting both the expressions into equation \eqref{eq:dynamic_general} yields the closed nonlinear dynamics of capillary rise in active fluids as,
\begin{align}
\zeta\beta\nu\dot h^2
+
\left[
\frac{8\eta}{a^2}h
+
\zeta(S_0\nu+\beta\xi_0)
+
\zeta(\alpha\nu+\beta\mu)h
\right]\dot h
\notag\\
+
\left[
\rho g+\zeta(S_0\mu+\alpha\xi_0)
\right]h
+
\zeta\alpha\mu h^2
+
\zeta S_0\xi_0
-
\frac{2\gamma\cos\theta}{a}
=
0 .
\label{eq:nonlinear_capillary_rise}
\end{align}

The constitutive relations adopted here correspond to a generic isotropic active fluid in the linear-response regime, where the activity coefficients depend weakly on both the system configuration and the rate of deformation. Such behaviour is commonly observed in active suspensions including bacteria \citep{chaudhuri2025mechanochemical}, active nematics \citep{doostmohammadi2018active} and actomyosin gels \citep{ramaswamy2010mechanics,marchetti2013hydrodynamics,adkins2022dynamics}.

\subsection{Equilibrium height}

Setting $\dot h=0$ in equation \eqref{eq:nonlinear_capillary_rise} yields the condition for the
equilibrium height of the liquid column during capillary rise of active fluids. The resulting algebraic equation is quadratic in $h$, and can be written in the form,
\begin{equation}
A_2 h^2 + A_1 h + A_0 = 0 .
\end{equation}
The expressions for the coefficients $A_2, A_1,$ and $A_0$ is given by, $\displaystyle A_2 = \zeta\alpha\mu, A_1 = \zeta(S_0\mu+\alpha\xi_0)+\rho g,$ and $\displaystyle A_0 = \zeta S_0\xi_0-\left({2\gamma\cos\theta}/{a}\right)$, respectively. Solving this equation gives the equilibrium height ($h_{\infty}^a$),
\begin{equation}\label{eq:quadratic}
h_{\infty}^a
=
\frac{-A_1 \pm \sqrt{A_1^2-4A_2A_0}}{2A_2}, \qquad (A_2\neq 0).
\end{equation}
In the special case where the coefficient of the quadratic term vanishes,
\emph{i.e.} $A_2=0$, the governing equation becomes linear in $h$.
The equilibrium height then reduces to,
\begin{equation}\label{eq:h_eq}
h_{\infty}^a
=
\frac{
\displaystyle \left({2\gamma\cos\theta}/{a}\right)-\zeta S_0\xi_0
}{
\rho g+\zeta(S_0\mu+\alpha\xi_0)
}.
\end{equation}

Asymptotically this reduces to the classical Jurin's law (equation~\ref{eq:jurin}) for passive fluids if $\zeta = 0$ is imposed in equation~\eqref{eq:quadratic} or~(\ref{eq:h_eq}).

\subsection{Inertial extension: active Bosanquet formulation}

The modified LW equation (\ref{eq:mod_LW}) describes the
viscous-dominated regime of capillary rise and neglects the inertia of the
liquid column. This approximation is valid at sufficiently long times when the
rise velocity becomes small. However, during the early stages of imbibition the
liquid column accelerates and inertial effects may become important.

To account for this regime we extend the dynamic balance to include the inertia
of the rising liquid column. The momentum of the column of height $h(t)$ and
cross-sectional area $\pi a^2$ is $\rho \pi a^2 h \dot h$, and the inertial
contribution therefore enters through the acceleration term
$\rho\,d(h\dot h)/dt$. The resulting dynamical equation becomes,

\begin{equation}
\rho \frac{d}{dt}(h\dot h)
+
\frac{8\eta}{a^2}h\dot h
=
\frac{2\gamma\cos\theta}{a}
-
\zeta S\left[(\mathbf{p}\cdot\mathbf{n})^2-\frac13\right]
-
\rho g h .
\label{eq:active_bosanquet}
\end{equation}

Equation (\ref{eq:active_bosanquet}) represents an \emph{active Bosanquet
equation}, which generalizes the classical inertial capillary rise model to
active fluids \citep{Bosanquet01031923}. The first term accounts for the inertia of the liquid column,
the second term corresponds to viscous resistance within the capillary tube,
while the terms on the right-hand side represent capillary driving,
activity-induced normal stress, and hydrostatic pressure respectively.

Using the identity, $\displaystyle \frac{d}{dt}(h\dot h)=\dot h^2+h\ddot h ,$ equation (\ref{eq:active_bosanquet}) may also be written in the form,

\begin{equation}
\rho(\dot h^2+h\ddot h)
+
\frac{8\eta}{a^2}h\dot h
=
\frac{2\gamma\cos\theta}{a}
-
\zeta S\xi
-
\rho g h ,
\end{equation}

where, $\xi=(\mathbf{p}\cdot\mathbf{n})^2-1/3$ denotes the interfacial alignment
factor introduced earlier.

The inertial contribution is dominant during the initial stages of capillary
rise, where the liquid column accelerates rapidly. As the rise proceeds, the
velocity decreases and viscous resistance becomes the dominant dissipation
mechanism. In this long-time limit, the inertial term becomes negligible and
equation (\ref{eq:active_bosanquet}) reduces to the modified Lucas--Washburn
equation (\ref{eq:mod_LW}). The equilibrium height obtained from equation 
(\ref{eq:active_bosanquet}) is identical to that derived previously, since the
inertial and viscous terms vanish when, $\dot h=0$. Consequently, the final
steady height is still governed by the generalized Jurin's law for active fluids (equation 
(\ref{gen_Jurin_law})), while inertia only affects the transient approach to
equilibrium.

\section{Results and discussion}

The mathematical formulation derived above reveals that activity modifies
capillary rise through the active normal stress contribution,
$-\zeta S\xi$ (equation~\ref{eq:dynamic_general}) appearing in the Young--Laplace condition. 
The resulting equilibrium height ($h_{\infty}^a$) therefore differs fundamentally from the
classical Jurin height ($h_J$) and depends on the orientational order of the active
constituents and their alignment with the interface.

\subsection{Active modification of Jurin's law}

For the simplest case where the orientational order parameter and the
alignment factor remain constant, \emph{i.e.}
$S=S_0$ and $\xi=\xi_0$, the equilibrium relation reduces to,
\begin{equation}\label{eq:active_jurin_law_1}
h_{\infty}^a
=
\frac{1}{\rho g}
\left(
\frac{2\gamma\cos\theta}{a}
-
\zeta S_0\xi_0
\right).
\end{equation}

This expression represents the simplest form of the
\emph{active Jurin's law}. 
The second term corresponds to an activity-induced normal stress that
either enhances or suppresses capillary rise depending on its sign. To make the competition between capillarity and activity explicit,
we introduce the classical Jurin height ($h_J$) from equation~\eqref{eq:jurin}. Using this scale, the equilibrium relation may be rewritten in dimensionless
form ($H_\infty = {h_{\infty}^a}/{h_J}$) as,
\begin{equation}\label{eq:active_jurin_law}
H_\infty = 1-
\mathrm{Ja}_a\,\xi_0 ,
\end{equation}

where,
$\displaystyle \mathrm{Ja}_a
=
\frac{\zeta S_0 a}{2\gamma\cos\theta},$
is an \emph{active Jurin number} measuring the ratio of the active
stress scale to the classical capillary pressure. For a fixed alignment state, the normalized rise height follows directly from equation 
(\ref{eq:active_jurin_law}). The sign and magnitude of the product
$\mathrm{Ja}_a\xi_0$ determine how activity modifies capillary rise.
The sign of the activity-induced correction therefore depends not only on
the sign of the activity coefficient $\zeta$, which distinguishes
contractile and extensile systems, but also on the interfacial alignment
factor $\xi_0$ that characterizes the orientation of active constituents
relative to the interface.

\subsection{Dynamic scaling of active capillary rise}

\begin{figure}
\centerline{\includegraphics[width=\linewidth]{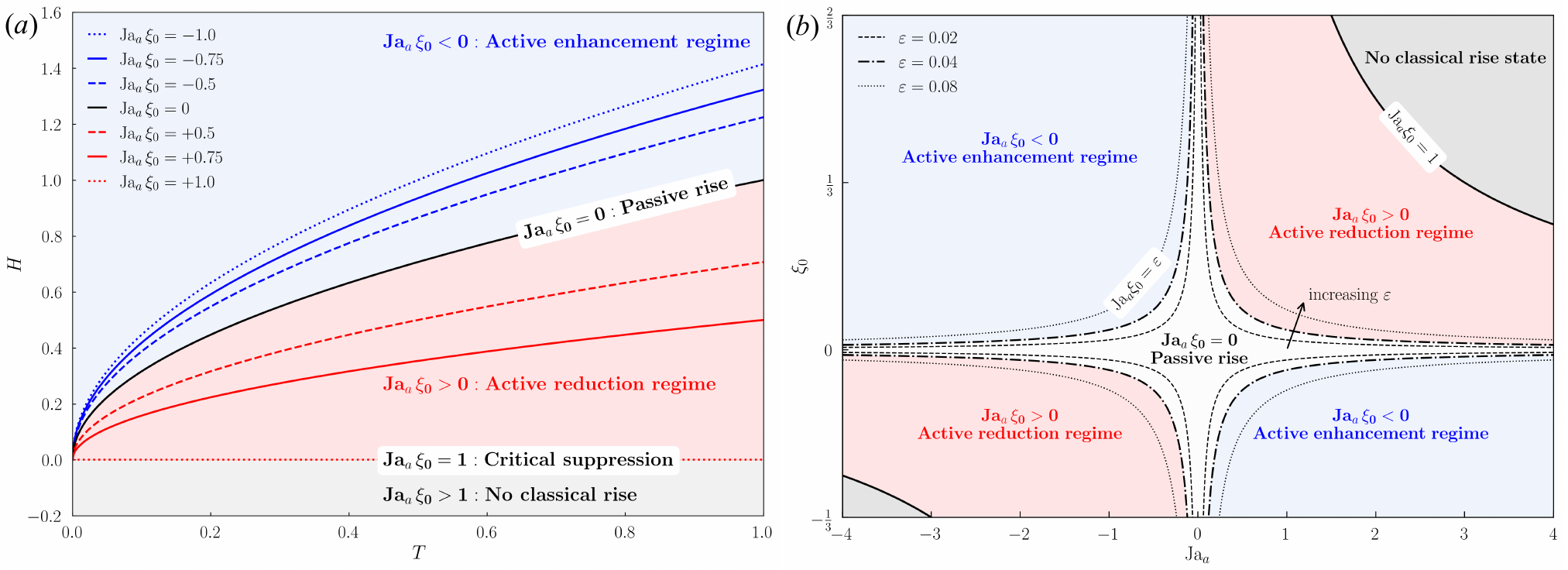}}
\caption{
($a$) Predicted rise dynamics obtained from equation (\ref{eq:LW_dimensionless})
for different values of the activity parameter $\mathrm{Ja}_a\xi_0$.
Activity enhances or suppresses the rise velocity while preserving
the classical $H\sim T^{1/2}$ scaling. ($b$) Regime map in ($Ja_a, \xi_0$) plane for active Jurin's law. 
}
\label{fig1}
\end{figure}

For the simplest case, where, $S=S_0$ and $\xi=\xi_0$ remain constant, and during the early stages of imbibition, the hydrostatic term is small ($h\ll h_J$), then, the equation (\ref{eq:dynamic_general}) reduces to,

\begin{equation}
\frac{8\eta}{a^2}h\dot h
=
\frac{2\gamma\cos\theta}{a}
-
\zeta S_0\xi_0 .
\end{equation}

Integrating this, we obtain the dynamics of the capillary rise as
\begin{equation}\label{eq:LW_scaling}
h^2
=
\frac{a}{4\eta}
\left(
2\gamma\cos\theta
-
\zeta S_0\xi_0 a
\right)t  
=
\frac{a\gamma\cos\theta}{2\eta}
\left(1-\mathrm{Ja}_a\xi_0\right)t.
\end{equation}

In order to express the result in dimensionless form, we introduce the classical Jurin height ($h_J$)
and the viscous capillary-rise timescale,
\(
t_J = {8\eta h_J}/{(a^2\rho g)}.
\)
Defining the dimensionless variables,
\(
H = h/h_J
\)
and
\(
T = t/t_J,
\)
equation \eqref{eq:LW_scaling} can be written as
\begin{equation}\label{eq:LW_dimensionless}
H^2 = (1-\mathrm{Ja}_a\xi_0)\,T.
\end{equation}

Thus the Lucas--Washburn scaling $H\sim T^{1/2}$ remains intact, while activity modifies the prefactor through the factor $(1-\mathrm{Ja}_a\xi_0)$. The rise is therefore accelerated for $\mathrm{Ja}_a\xi_0<0$ and slowed for $\mathrm{Ja}_a\xi_0>0$, as shown in figure~\ref{fig1}.

\subsection{Regimes of active capillarity}

Equation~\eqref{eq:active_jurin_law} shows that capillary rise in an active fluid is controlled by the competition between the classical Laplace pressure and the activity-induced normal stress at the meniscus. 
This competition is quantified by the dimensionless activity parameter, $\mathrm{Ja}_a\xi_0$. Depending on the magnitude and sign of $\mathrm{Ja}_a\xi_0$, the activity-induced stress may either augment, oppose, or completely overwhelm the capillary suction driving the rise. Consequently, several distinct regimes of active capillarity emerge.

When $|\mathrm{Ja}_a\xi_0|\ll 1,$ equation \eqref{eq:active_jurin_law} reduces to,
$H_\infty
=
1-\mathrm{Ja}_a\xi_0
=
1+O(\mathrm{Ja}_a\xi_0).$ In this regime, activity acts only as a weak perturbation to the classical Jurin height. 
The meniscus remains capillarity-dominated, and the active fluid behaves effectively as a \emph{passive fluid} with a weakly renormalized capillary pressure. This regime corresponds to the classical \emph{passive-fluid limit}. For the purpose of constructing the regime map, this asymptotic condition is quantified by introducing a small cutoff parameter, $\varepsilon\ll1$ and defining the passive-like regime as, 
$|\mathrm{Ja}_a\xi_0|\le\varepsilon $. The curves $|\mathrm{Ja}_a{\xi}_0|=\varepsilon$ therefore mark the boundaries of the passive limit in figure~\ref{fig1}($b$). 
Different choices of $\varepsilon$ simply reflect how strictly the condition $|\mathrm{Ja}_a\xi_0|\ll1$ is enforced.

When $|\mathrm{Ja}_a\xi_0|=O(1),$ activity contributes at the same order as the Laplace pressure. 
The equilibrium height is then strongly modified according to, 
$H_\infty = 1-\mathrm{Ja}_a\xi_0.$ If $\mathrm{Ja}_a\xi_0<0$, the rise height exceeds the classical Jurin height, whereas if $\mathrm{Ja}_a\xi_0>0$, the height is reduced below $h_J$. This regime corresponds to an active renormalization of capillarity and is the simplest experimentally accessible signature of the theory which can be termed as \emph{active enhancement or reduction regime} as shown in figure~\ref{fig1}.

A particularly interesting transition occurs when $\mathrm{Ja}_a\xi_0=1$ (see figure~\ref{fig1}). At this transition point, equation~\eqref{eq:active_jurin_law} yields, $H_\infty=0$. Thus the capillary rise state is completely suppressed even though the substrate may remain wetting leading to \emph{critical suppression}. For $\mathrm{Ja}_a\xi_0>1,$ the formal expression gives $H_\infty<0$, which indicates that no upward capillary rise is possible within the present static theory. 
Physically, this means that the activity-induced normal stress acts as an anti-capillary pressure strong enough to cancel and then overcome the Laplace suction. 
In this regime, activity effectively turns a wetting capillary into a non-rising one, making this an \emph{active suppression regime}.

The most striking consequence of equation \eqref{eq:active_jurin_law} is that activity can drive imbibition even when the surface is non-wetting. 
For a non-wetting substrate, $\cos\theta<0$, so the classical Jurin height ($h_J$) is negative and passive capillary rise does not occur. 
Returning to the dimensional balance (equation \eqref{eq:active_jurin_law_1}), a positive rise becomes possible whenever, $\displaystyle -\zeta S_0\xi_0 > \left|{2\gamma\cos\theta}/{a}\right|.$ Under this condition, the active stress generates an effective suction pressure that exceeds the adverse capillary pressure of the non-wetting meniscus. 
The liquid is then drawn upward not by capillarity, but by internally generated active stress. 
This phenomenon may be interpreted as \emph{activity-driven imbibition} or \emph{active pumping} by the meniscus on \emph{non-wetting surfaces}.

The constant-$S_0$ result in equation~\eqref{eq:active_jurin_law} is only the leading-order description. 
If the orientational order and alignment evolve with height, the active correction becomes state dependent, as shown in equation~\eqref{eq:quadratic}. When the discriminant is positive, $A_1^2-4A_2A_0>0$, equation~\eqref{eq:quadratic}
admits two distinct equilibrium heights. This suggests the presence of multiple
steady states, which may lead to \emph{activity-induced bistability} provided both
solutions are positive and stable to small perturbations.
Such behaviour has no passive analogue in the simplest Jurin problem and arises entirely from the nonlinear feedback between capillary confinement and active orientational order. This limit corresponds to the probable \emph{activity-induced bistable regime}.

To distinguish the existence of multiple equilibrium heights from genuine bistability, we perform LSA of the capillary-rise dynamics. In the overdamped limit the evolution equation may be written in the form $B(h)\dot h + P(h)=0$, where $P(h)=A_2h^2+A_1h+A_0$ and $\displaystyle B(h) = ({8\eta}/{a^2})h+\zeta(S_0\nu+\beta\xi_0)+
\zeta(\alpha\nu+\beta\mu)h > 0$ is an effective dissipative coefficient. An equilibrium height, $h=h^*$ satisfies $P(h^*)=0$. To assess its stability we introduce a small perturbation $h(t)=h^*+\delta h(t)$ with $|\delta h|\ll1$ and linearize the dynamics about $h^*$. Retaining terms to first order in $\delta h$ gives, $B(h^*)\,\delta\dot h + P'(h^*)\,\delta h = 0,$ which admits normal-mode solutions, $\delta h\sim e^{\omega t}$ with growth rate, $\omega=-P'(h^*)/B(h^*)$. The sign of $\omega$ therefore determines stability. Since $P'(h)=2A_2h+A_1$, the two solutions of $P(h)=0$ satisfy $P'(h_+)=-P'(h_-)$ and hence are generically of opposite stability when $B(h^*)>0$. The overdamped dynamics therefore admits multiple steady states but not true bistability.

The situation changes when inertia is retained, as in the \emph{active Bosanquet formulation} introduced in \S2.3. The governing equation then takes the second-order form, $\rho h\,\ddot{h} + B(h)\dot{h} + P(h) = 0,$ which may be interpreted as the motion of a damped particle with height-dependent inertia in an effective potential $U(h)$ defined through $U'(h)=P(h)$. In this formulation the equilibrium heights satisfy $P(h_*)=0$ and therefore correspond to the extrema of $U(h)$. Linearization about an equilibrium shows that it is dynamically stable when $P'(h_*)>0$, corresponding to a local minimum of $U(h)$, and unstable when $P'(h_*)<0$, corresponding to a local maximum. If activity renders the effective force $P(h)$ non-monotonic, the resulting potential may become non-convex and develop two minima separated by a maximum. In that case the system admits two dynamically stable rise heights and one unstable intermediate state. The capillary column can therefore relax to different final heights depending on the initial condition of the meniscus, signalling bistability. The inertial extension thus provides the natural dynamical framework in which such \emph{activity-induced bistability} can emerge in the active fluid capillary-rise problem.

If the orientational order and alignment also depend on the rise speed, the capillary dynamics becomes intrinsically nonlinear, as discussed in \S\ref{linear_response_model}. The flow dependence therefore acts not only on the effective capillary driving but also on the effective hydrodynamic resistance. 
Such coupling can generate overshoot, strong slowing down, or other nonlinear rise trajectories even in a simple cylindrical capillary. 
In this sense, the active capillary problem goes beyond a static renormalization of Jurin's law and becomes a genuinely nonlinear active transport problem.

The different capillary behaviours predicted by the theory may be summarized as follows: 
($a$) when $|\mathrm{Ja}_a\xi_0|\ll 1$, activity acts only as a weak perturbation and the system behaves in a passive-like capillary regime;
($b$) when $\mathrm{Ja}_a\xi_0<0$, the activity-induced stress augments the Laplace pressure, leading to activity-enhanced capillary rise;
($c$) for $0<\mathrm{Ja}_a\xi_0<1$, activity opposes the capillary suction and reduces the equilibrium height; 
($d$) at the critical condition, $\mathrm{Ja}_a\xi_0=1$, the active stress exactly balances the Laplace pressure and capillary rise is completely suppressed; 
($e$) when $\mathrm{Ja}_a\xi_0>1$, the active stress overwhelms capillarity and no classical rising state exists; 
($f$) if $-\zeta S_0\xi_0>\left|2\gamma\cos\theta/a\right|$, activity can drive imbibition even on nominally non-wetting surfaces; and 
($g$) finally, when the nonlinear condition $A_1^2>4A_2A_0$ is satisfied, the inertial system admits two steady heights, which may give rise to \emph{activity-induced bistability}. The active Jurin's law, equation~\eqref{eq:active_jurin_law} therefore provides a compact framework for understanding how internally generated stresses reshape one of the most classical capillary problems in fluid mechanics.

\section{Conclusions}

We have developed a generalized theory of capillary rise in active fluids by incorporating activity-induced stresses into the interfacial normal stress balance. The resulting framework extends the classical Jurin's law by introducing an additional pressure contribution arising from internally generated active stresses. This modification leads to an activity-dependent equilibrium height and altered capillary rise dynamics. In particular, extensile and contractile systems can produce qualitatively different behaviours, leading to either enhancement or suppression of capillary rise depending on the sign and magnitude of the activity and the interfacial alignment of the active constituents.

A central outcome of the analysis is the emergence of an \emph{active Jurin number}, which quantifies the competition between internally generated active stresses and the classical capillary pressure. This dimensionless number and the alignment factor provides a simple criterion for predicting whether activity enhances, reduces or completely suppresses capillary rise. In contrast to passive fluids, the theory predicts regimes where capillary rise may vanish entirely, and where imbibition can occur even on nominally non-wetting surfaces when the active stress exceeds the Laplace pressure. Such behaviour has no analogue in classical capillarity and arises solely from the internally generated stresses of active fluids. The analysis also reveals that nonlinear coupling between orientational order and capillary dynamics can lead to multiple equilibrium heights. While LSA of the overdamped dynamics admits multiple equilibrium heights of opposite stability, and therefore no true bistability, the inertial extension provided by the active Bosanquet formulation introduces second-order dynamics that may produce two stable rise heights separated by an unstable state, offering a route to genuine activity-induced bistability depending on the system parameters.

These predictions suggest that capillary rise experiments in microcapillaries containing bacterial suspensions, active nematics or cytoskeletal extracts may provide a simple route to probing active stresses and interfacial alignment in active fluids. More broadly, the present framework highlights capillary rise as a simple yet powerful probe of active matter systems, and suggests that extensions including interfacial elasticity, activity-driven Marangoni stresses or deformable interfaces may reveal further ways in which active stresses reshape classical interfacial transport phenomena.




\begin{acknowledgments}
J. C. acknowledges the financial support provided by SRIC of the Indian Institute of Technology Kharagpur, India, through the Faculty Start-up Research Grant (FSRG) (IIT/SRIC/CH/SON/2025-2026/227) scheme. The authors have no conflicts of interest to disclose.
\end{acknowledgments}

\section*{Author Contributions}
The authors jointly designed the study and analysed the results.


\bibliographystyle{jfm}
\bibliography{manuscript}

\end{document}